\documentclass
[superscriptaddress,secnumarabic,amssymb,amsmath,nobibnotes,aps,prd,showkeys,showpacs,nofootinbib,onecolumn]{revtex4}%
\usepackage{graphicx}
\usepackage{subfigure}
\usepackage{epstopdf}
\usepackage{epsf}
\usepackage{bm}
\usepackage{amsmath}
\usepackage{amsfonts}
\usepackage{amssymb}%
\providecommand{\U}[1]{\protect\rule{.1in}{.1in}}

\newcommand{\be}{\begin{equation}}
\newcommand{\ee}{\end{equation}}

\newcommand{\mincir}{\raise
-3.truept\hbox{\rlap{\hbox{$\sim$}}\raise4.truept\hbox{$<$}\ }}
\newcommand{\magcir}{\raise
-3.truept\hbox{\rlap{\hbox{$\sim$}}\raise4.truept\hbox{$>$}\ }}

\begin{document}
\title{Asymptotic behavior of $N$-fields Chiral Cosmology}
\author{Andronikos Paliathanasis}
\email{anpaliat@phys.uoa.gr}
\affiliation{Institute of Systems Science, Durban University of Technology, Durban 4000,
South Africa}
\affiliation{Instituto de Ciencias F\'{\i}sicas y Matem\'{a}ticas, Universidad Austral de
Chile, Valdivia 5090000, Chile}
\author{Genly Leon}
\email{genly.leon@ucn.cl}
\affiliation{Departamento de Matem\'{a}ticas, Universidad Cat\'{o}lica del Norte, Avda.
Angamos 0610, Casilla 1280 Antofagasta, Chile.}

\begin{abstract}
We perform a detailed analysis for the asymptotic behaviour for the
multi-scalar field Chiral cosmological scenario. We present the asymptotic
behaviour for the one-field, two-fields and three-fields Chiral models. From
these results, and deriving conserved quantities,  we present a Theorem for the $N$-fields model for  the  Chiral  model  with $N$--fields. We find
that the maximum number of scalar fields which provide interesting physical
results is two-fields, while for $N>2$ the new stationary points are only of
mathematical interest since they do not describe new exact solutions
different from those recovered for $N=2$. 

\end{abstract}
\date{\today}
\maketitle

\section{Introduction}

\label{sec1}

The detailed analysis of the recent cosmological observations indicated that
our universe has  gone through two acceleration phases
\cite{dataacc1,dataacc2,data1,data2,Hinshaw:2012aka,Ade:2015xua,Aghanim:2018eyx}%
. A late-time acceleration phase \cite{ss1} and an early acceleration
phase known as inflation \cite{Aref1,guth}. The main mechanics for the
description of the inflation era  is based on  a scalar
field, called inflaton \cite{ref2,ref3,newinf,ref4}, which dominates in the early universe and drives into the current acceleration epoch. Moreover, the
cause of the late-time acceleration is also unknown. The late-time
acceleration it is attributed to a matter source called Dark-Energy. 

Dark energy would need to have a strong negative pressure (repulsive action), to explain the observed acceleration of the expansion of the universe.  In the Friedmann–Lemaître–Robertson–Walker metric, it can be shown that a strong constant negative pressure in all the universe causes an acceleration in the expansion if the universe is already expanding, or a deceleration in contraction if the universe is already contracting. This accelerating expansion effect is sometimes labeled ``gravitational repulsion''. There are various attempts by the cosmologists to determine the physics behind
the Dark Energy. The main approaches are based on the introduction of scalar
fields in the gravitational field equations \cite{Ratra,Barrow,Linder,si1,si4},  or are based on the introduction of other geometrodynamical terms which follow
from the modification of the Einstein-Hilbert Action\ \cite{cl1,cl2}. In the
latter modified theories of gravity, the geometrodynamical terms can be described by
scalar fields by attributing the new degrees of freedom. For an extended
discussion we refer the reader to \cite{sotsf}.

The simplest scalar field which has been introduced in the cosmological
studies is the quintessence field, \cite{Ratra}, next it was introduced
the phantom field or the k-essence theory \cite{si1,si4}. Other Scalar field
models which have been proposed in the literature are nonminimally coupled
with gravity, such as the Brans-Dicke theory \cite{Brans}, which belongs to
the Scalar tensor theories \cite{Hanlon,faraonibook}, or the more general
Hordenski Action Integral \cite{hor} or its special cases like Galileon or
cubic gravity \cite{gal02,nik,Leon:2012mt,DeArcia:2015ztd,cub,DeArcia:2018pjp}. Furthermore, it was introduced  the teleparallel analog of Horndeski gravity \cite{Bahamonde:2019shr,Bahamonde:2019ipm,Bahamonde:2020cfv}, and $f(T,T_G)$ class of gravitational modification, based on the quadratic torsion scalar $T$, as well as on the new quartic torsion scalar $T_G$ which is the teleparallel equivalent
of the Gauss-Bonnet term \cite{Kofinas:2014owa,Kofinas:2014aka}. 

In order to overpass various problems in the description of the
universe, multi-field cosmological models have been proposed. For instance, the two scalar field model known as quintom  model
consists of a quintessence and one phantom field \cite{qq1}. In the quintom
model, the parameter for the equation of state of the effective cosmological
fluid can cross the phantom divide line more than once \cite{qq1,qq2,Lazkoz:2006pa,qq3,Leon:2012vt,Leon:2018lnd},
without the existence of ghosts. Furthermore, multi-field models have been
proposed to describe  different mechanisms for the inflationary era
\cite{hy4,atr1,atr3}. Other multi-field models have been used to describe the
``dark sector'' of the universe, that is, the dark energy and the dark
matter \cite{mm1,mm2,mm3,mm4,mm5}. In \cite{cc1} the dynamics of a
multi-scalar field cosmology model \ was studied for a model which consists of
 various quintessence fields with independent scalar field
potential.\ In this study, there was found that the model provides a new dynamical
behaviour with physical importance. However, a questions which follows is:
which is the number of scalars fields in the multi-field theory which provides results of physical interests.

In this work, we are interested in the multi-field theory known as Chiral
model \cite{atr7,sigm0}. In particular we study the asymptotic behaviour of
this multi-field model and we attempt to answer  the questions of what the
maximum number of fields is, and when we get the results of physical
interests from. The model of our consideration is the $N-$field model proposed in
\cite{nfield} and generalizes the two-scalar Chiral model \cite{ns01}. In
particular, the Lagrangian function for the $N$-fields  is considered to
have a nonlinear kinetic part, such that the kinetic quantities of the fields
would define the $N$-dimensional manifold with a maximally symmetric line element
of constant nonzero curvature. For this specific theory and for various
families of scalar field potentials, exact and analytic solutions have been
determined \cite{nfield}. The plan of the paper is as follows.

In Section \ref{sec2} ,we present the two-scalar field Chiral model where we
discuss the main properties of the model and we show how the fields can
interact. The $N-$field Chiral
model  is discussed in Section \ref{sec3}. The study for asymptotic
behaviour of the cosmological field equations for a spatially flat FLRW
background space is given in Section \ref{sec4}. More precisely, we present in
detail the analysis for the asymptotic behaviour for the one-scalar field
model, that is, the quintessence model, the two-scalar fields Chiral model,
the three-scalar fields Chiral model, from where we observe that when we pass
from one to  two fields, new stationary points which describe different
eras in the cosmological history exist. Moreover, when we pass from the
two-field to the three-field model, we observe that new stationary points 
are obtained, but without new physical interests. In the two- and three-scalar field
models, the only possible attractors are those of the quintessence theory,
which  is an interesting result because these multi-field models can be
used either to describe the matter era if the model will be considered as
a unification of the dark-energy and dark matter model, or by being considered as
an inflationary model with two different behaviours in the inflationary era.
As far as the $N-$fields model is concerned, the asymptotic behaviour is
described by a theorem, from where we conclude that the main points of
special physical interests are found for $N=2$, and the models with $N>2$ have only
mathematical interests. Finally, in Section \ref{sec5}, we draw our conclusions.

\section{Chiral cosmology}

\label{sec2}

In Chiral model the gravitational Action Integral  consists of  two scalar
fields \cite{ns01}%
\begin{equation}
S=\int dx^{4}\sqrt{-g}\left(  R-\frac{1}{2}g^{\mu\nu}\nabla_{\mu}\phi
\nabla_{\nu}\phi-\frac{1}{2}g^{\mu\nu}e^{2\sigma\phi}\nabla_{\mu}\psi
\nabla_{\nu}\psi-V\left(  \phi,\psi\right)  \right)  \label{ch.01}%
\end{equation}
where $g_{\mu\nu}$ is the metric tensor of the underlying four-dimensional
spacetime with Ricciscalar $R$ and $\phi=\phi\left(  x^{\mu}\right)  $,
$\psi=\psi\left(  x^{\mu}\right)  $ are the two scalar fields minimally
coupled to gravity but with interaction terms which depend on the coupling
constant $\sigma$, while function $V\left(  \phi,\psi\right)  $ is the scalar
field potential. We observe that the kinetic part of the two fields $\left\{
\phi,\psi\right\}  $ is defined by a space of constant curvature.

The Action Integral (\ref{ch.01}) can be seen as the generalization of a
complex scalar field model with Action Integral%
\begin{equation}
S=\int dx^{4}\sqrt{-g}\left(  R-\frac{1}{2}g^{\mu\nu}\nabla_{\mu}\Phi
\nabla_{\nu}\Phi^{\ast}-V\left(  \Phi,\Phi^{\ast}\right)  \right)
\label{ch.02}%
\end{equation}
in which $\Phi=\phi+i\psi~$and the norm $\left\vert \Phi\right\vert ^{2}%
=\Phi\Phi^{\ast}$ is defined as
\begin{equation}
\left\vert \Phi\right\vert ^{2}=\phi^{2}+e^{2\sigma\phi}\psi^{2}.
\label{ch.03}%
\end{equation}

For the background space we consider a spatially flat FLRW spacetime which is
described by the line element%
\begin{equation}
ds^{2}=-dt^{2}+a^{2}\left(  t\right)  \left(  dx^{2}+dy^{2}+dz^{2}\right)  ,
\label{ch.04}%
\end{equation}
where $a\left(  t\right)  $ is the scale factor and $H=\frac{\dot{a}}{a}$ is
the Hubble function. The FLRW spacetime admits a six dimensional Killing
algebra, which we assume that it is inherited by the two scalar
fields~$\left\{  \phi,\psi\right\}  $, that is $\phi=\phi\left(  t\right)  $
and $\psi=\psi\left(  t\right)  $.

By replacing the Ricciscalar of the line element (\ref{ch.04}) in
(\ref{ch.01}) the gravitational field equations follows \cite{ns02}
\begin{equation}
3H^{2}=\frac{1}{2}\left(  \dot{\phi}^{2}+e^{2\sigma\phi}\dot{\psi}^{2}\right)
+V\left(  \phi,\psi\right)  ,\label{ch.05}%
\end{equation}%
\begin{equation}
2\dot{H}+3H^{2}=-\left(  \frac{1}{2}\left(  \dot{\phi}^{2}+e^{2\sigma\phi}%
\dot{\psi}^{2}\right)  -V\left(  \phi,\psi\right)  \right)  ,\label{ch.06}%
\end{equation}
while the scalar fields satisfy the continuity equations%
\begin{equation}
\ddot{\phi}+3H\dot{\phi}-\sigma e^{2\sigma\phi}\dot{\psi}^{2}+V_{,\phi
}=0,\label{ch.07}%
\end{equation}%
\begin{equation}
\ddot{\psi}+3H\dot{\psi}+2\sigma\dot{\phi}\dot{\psi}+e^{-2\sigma\phi}V_{,\psi
}=0.\label{ch.08}%
\end{equation}

The field equations (\ref{ch.05})-(\ref{ch.08}) have been widely studied in
the literature. Some exact solutions are presented in \cite{ns01} while
functional forms for the scalar field potential for which the field equations form
a Liouville integrable system were determined in \cite{ns02}. Scaling
attractors in Chiral model were found in \cite{ns03} and a detailed analysis
of the dynamics of Chiral model was presented in \cite{ns04}. Moreover in
\cite{ns05} extensions of the Chiral\ Action Integral (\ref{ch.01}) were studied.

However this is not the only parametrization for the Chiral model. Indeed
without loss of generality we can define the new fields $\Phi,~\Psi$ such that
the Action Integral to be \cite{ns02}
\begin{equation}
S=\int dx^{4}\sqrt{-g}\left(  R-\frac{1}{2}g^{\mu\nu}\nabla_{\mu}\Phi
\nabla_{\nu}\Phi-\frac{1}{2}g^{\mu\nu}\sinh^{2}\left(  \kappa\Phi\right)
\nabla_{\mu}\Psi\nabla_{\nu}\Psi-V\left(  \Phi,\Psi\right)  \right)
.\label{ch.09}%
\end{equation}

There is a one to one relation between the fields $\left\{  \phi,\psi\right\}
\rightarrow\left\{  \Phi,\Psi\right\}  $, the two Action integrals
(\ref{ch.01}), (\ref{ch.09}) are related under the action of a point
transformation on the space of constant curvature for the scalar fields. In
the following we are interested on the parametrization of the Action Integral
(\ref{ch.09}).

For mass-free scalar fields $\left\{  \Phi,\Psi\right\} $, and in the presence
of the cosmological constant term,  it was found in \cite{nsdim} that the
equation of state parameter for the effective fluid can cross the phantom
divide line as an effect of quantum transitions.

\section{$N-$field Chiral model}

\label{sec3}

A multi-field generalization of the Chiral model it was proposed in
\cite{nfield}. In particular we consider the multi-field Action Integral
\cite{nfield}%
\begin{equation}
S=\int\sqrt{-g}dx^{4}R-\int\sqrt{-g}dx^{4}L_{\Phi}\left(  \Phi^{C},\nabla
_{\mu}\Phi^{C}\right), \label{ch.10}%
\end{equation}
where $L_{\Phi}\left(  \Phi^{C},\nabla_{\mu}\Phi^{C}\right)  $ is the Action
integral for the $N$-fields $\Phi^{A}=\left(  \phi_{1},\phi_{2},...,\phi
_{N}\right)  $ given as follows \cite{nfield}%
\begin{equation}
L_{\Phi}\left(  \Phi^{C},\nabla_{\mu}\Phi^{C}\right)  =\frac{1}{2}g^{\mu\nu
}\Xi_{AB}\left(  \Phi^{C}\right)  \nabla_{\mu}\Phi^{A}\nabla_{\nu}\Phi
^{B}+V\left(  \Phi^{C}\right),  \label{ch.11}%
\end{equation}
were $\Xi_{AB}\left(  \Phi^{C}\right)  $ is a second-rank tensor and defines
the space where the scalar fields evolves. For the quintom model, $\Xi
_{AB}\left(  \Phi^{C}\right)  $ is a two-dimensional flat space of Lorentzian
signature, while for the Chiral model $\Xi_{AB}\left(  \Phi^{C}\right)  $ is a
two-dimensional space of constant (negative) curvature, for instance
\begin{equation}
\Xi_{AB}\left(  \Phi^{C}\right)  =diag\left(  1,\sinh^{2}\left(  \kappa
\phi_{1}\right)  \right)  .\label{ch.12}%
\end{equation}

Therefore, in order to consider a $N-$field extension of the Chiral model we
assume that $\Xi_{AB}\left(  \Phi^{C}\right)  $ is a $N$-dimensional space of
constant curvature, that is, \cite{nfield}%
\begin{equation}
\Xi_{AB}\left(  \Phi^{C}\right)  =diag\left(  1,\sinh^{2}\left(  \kappa
\phi_{1}\right)  ,\sinh^{2}\left(  \kappa\phi_{1}\right)  \sin^{2}\left(
\kappa\phi_{2}\right)  ,...,\sinh^{2}\left(  \kappa\phi_{1}\right)  \sin
^{2}\left(  \kappa\phi_{2}\right)  ...\sin^{2}\left(  \kappa\phi_{N-1}\right)
\right)  .\label{ch.13}%
\end{equation}

For the line element (\ref{ch.10}) with Lagrangian density (\ref{ch.11}) the
gravitational field equations are%
\begin{equation}
G_{\mu\nu}=T_{\mu\nu}\left(  \Phi^{C},\nabla_{\mu}\Phi^{C}\right)  ,
\label{ch.14}%
\end{equation}
in which%
\begin{equation}
T_{\mu\nu}\left(  \Phi^{C},\nabla_{\mu}\Phi^{C}\right)  =\Xi_{AB}\left(
\Phi^{C}\right)  \nabla_{\mu}\Phi^{A}\nabla_{\nu}\Phi^{B}-g_{\mu\nu}\left(
-\frac{1}{2}g^{\mu\nu}\Xi_{AB}\left(  \Phi^{C}\right)  \nabla_{\mu}\Phi
^{A}\nabla_{\nu}\Phi^{B}-V\left(  \Phi^{C}\right)  \right)  , \label{ch.15}%
\end{equation}
while the equation of motions for the scalar fields are given by the
$N$-dimensional vector field%
\begin{equation}
g^{\mu\nu}\left(  \nabla_{\mu}\Xi_{~B}^{A}\left(  \Phi^{C}\right)  \nabla
_{\nu}\Phi^{B}\right)  +\Xi_{~B}^{A}\left(  \Phi^{C}\right)  \frac{\partial
V\left(  \Phi^{C}\right)  }{\partial\Phi^{B}}=0. \label{ch.16}%
\end{equation}

For the cosmological line element \eqref{ch.04} the Friedmann equations reads%
\begin{equation}
-3H^{2}+\frac{1}{2}\Xi_{AB}\left(  \Phi^{C}\right)  \dot{\Phi}^{A}\dot{\Phi
}^{B}+V\left(  \Phi^{C}\right)  =0, \label{ch.17}%
\end{equation}%
\begin{equation}
-2\dot{H}-3H^{2}+\frac{1}{2}\Xi_{AB}\left(  \Phi^{C}\right)  \dot{\Phi}%
^{A}\dot{\Phi}^{B}-V\left(  \Phi^{C}\right)  =0, \label{ch.18}%
\end{equation}
and the conservation laws are
\begin{equation}
\ddot{\Phi}^{A}+\tilde{\Gamma}_{BC}^{A}\left(  \Phi^{D}\right)  \dot{\Phi}%
^{B}\dot{\Phi}^{C}+3H\dot{\Phi}^{A}+\Xi^{AB}\left(  \Phi^{C}\right)
V_{,B}\left(  \Phi^{C}\right)  =0. \label{ch.19}%
\end{equation}
where $\tilde{\Gamma}_{BC}^{A}\left(  \Phi^{D}\right)  $ denotes the
Levi-Civita connection coefficients of $\Xi_{AB}\left(  \Phi^{C}\right)  $.

\section{Asymptotic behaviour}

\label{sec4}

In this section we study the effects of the additional scalar fields in the
asymptotic behaviour of the cosmological solution. In order to perform this
analysis we start our presentation by assuming one scalar field model
$\Phi^{A}=\phi$, which correspond to the quintessence model. For the scalar
field potential we consider the exponential potential $V\left(  \Phi^A\right)
=V_{0}e^{\lambda\phi}$. 

In order to study the asymptotic behaviour of the multi-field model, we define the dimensionless
variables \cite{copeland1}
\begin{align*}
~x_{1} &  =\frac{\dot{\phi}_{1}}{\sqrt{6}H}~~,\\
x_{2} &  =\sinh\left(  \kappa\phi_{1}\right)  \frac{\dot{\phi}_{2}}{\sqrt{6}%
H},\\
x_{3} &  =\sinh\left(  \kappa\phi_{1}\right)  \sin\left(  \kappa\phi
_{2}\right)  \frac{\dot{\phi}_{3}}{\sqrt{6}H},\\
&  \ldots \\
x_{N} &  =\sinh\left(  \kappa\phi_{1}\right)  \sin\left(  \kappa\phi
_{2}\right)  \ldots \sin\left(  \kappa\phi_{N-1}\right)  \frac{\dot{\phi}_{N}%
}{\sqrt{6}H}.
\end{align*}
and%
\[
y=\sqrt{\frac{V\left(  \phi_{1}\right)  }{3H^{2}}}~,~
\]
where the Friedmann equation reads%
\begin{equation}
1-y^{2}-\left(  \left(  x_{1}\right)  ^{2}+\left(  x_{2}\right)
^{2}+...+\left(  x_{N}\right)  ^{2}\right)  =0\label{ch.20}%
\end{equation}
while the equation of state parameter for the effective fluid becomes%
\begin{equation}
w_{eff}=\left(  x_{1}\right)  ^{2}+\left(  x_{2}\right)  ^{2}+...+\left(
x_{N}\right)  ^{2}-y^{2}.\label{ch.21}%
\end{equation}

\subsection{Quintessence}

For the quintessence model, that is, for one scalar field model, the
gravitational field equations are written in the form of the following system
\cite{copeland1}%
\begin{equation}
\frac{dx_{1}}{d\tau}=\frac{1}{2}\left(  3x_{1}^{3}-3x_{1}\left(
1+y^{2}\right)  -\sqrt{6}\lambda y^{2}\right)  ,\label{ch.22}%
\end{equation}%
\begin{equation}
\frac{dy}{d\tau}=\frac{1}{2}y\left(  3\left(  1+x_{1}^{2}-y^{2}\right)
+\sqrt{6}\lambda x_{1}\right)  .\label{ch.23}%
\end{equation}
where we have selected the new indepedent variable to be $\tau=\ln a$.

The stationary points $P=\left(  x_{1},y\right)  ~$of the system
(\ref{ch.22}), (\ref{ch.23}) which satisfy the constraint (\ref{ch.20}) are
\cite{copeland1}%
\[
\,P_{1}^{\left(  \pm\right)  }=\left(  \pm1,0\right)  ~,~P_{2}=\left(
-\frac{\lambda}{\sqrt{6}},\sqrt{1-\frac{\lambda^{2}}{6}}\right)  .
\]

The parameter for the equation of state for the effective fluid at the
stationary points are $w_{eff}\left(  P_{1}^{\left(  \pm\right)  }\right)  =1$
and $w_{eff}\left(  P_{2}\right)  =-1+\frac{\lambda^{2}}{3}$. The exact
solutions at the stationary points $P_{1}^{\left(  \pm\right)  }$ describe
universes where only the kinetic part of the scalar field dominates and the
exact solution is that of the stiff fluid source. On the other hand, at the
stationary point $P_{2}$ the scale factor describes a scaling solution and
acceleration occurs when $\left\vert \lambda\right\vert <\sqrt{2}$. Point
$P_{2}$ describes a de Sitter universe only when $\lambda=0$, which means that
the scalar field potential plays the role of the cosmological constant, whole
point $P_{2}$ is physically accepted when $\left\vert \lambda\right\vert
<\sqrt{6}$.

In order to study the stability of the stationary points we use the constraint
equation (\ref{ch.20}) to reduce the dynamical system (\ref{ch.22}),
(\ref{ch.23}) into the one dimensional equation \cite{copeland1}%
\[
\frac{dx_{1}}{d\tau}=3\left(  x^{2}-1\right)  \left(  x+\frac{\lambda}%
{\sqrt{6}}\right)
\]
from where we find that eigenvalues the linearized system at the stationary
points are
\[
e\left(  P_{1}^{\pm}\right)  =\sqrt{6}\lambda\pm6~~,~~e\left(  P_{2}\right)
=\frac{1}{2}\left(  \lambda^{2}-6\right)  .
\]
We conclude that point $P_{1}^{+}$ is an attractor when $\lambda<-\sqrt{6}$,
point $P_{1}^{-}$ is an attractor when $\lambda>\sqrt{6}$ while $P_{2}$ is
always an attractor \cite{copeland1}. We continue with the two-scalar field model.

\subsection{Two-field model}

For the two scalar fields model, $\Phi^{A}=\left(  \phi_{1},\phi_{2}\right)$, we consider the exponential dependence on the first scalar field, $V\left(  \Phi\right)  =V_{0}e^{\lambda\phi_{1}}$, for simplicity. Notice there is a nontrivial interaction between the scalar fields which follows from the nonlinear terms of the scalar field Lagrangian function.

The field equations in the dimensionless variables read%
\begin{align}
\frac{dx_{1}}{d\tau}  &  =\frac{1}{2}\left(  3x_{1}^{3}-3x_{1}\left(
1+y^{2}-x_{2}^{2}\right)  -\sqrt{6}\lambda y^{2}+\sqrt{6}\mu_{1}x_{2}%
^{2}\right)  ,\label{ch.24}\\
\frac{dx_{2}}{d\tau}  &  =\frac{1}{2}x_{2}\left(  3\left(  x_{1}^{2}+x_{2}%
^{2}-y^{2}-1\right)  -\sqrt{6}\mu_{1}x_{1}\right)  ,\label{ch.25}\\
\frac{dy}{d\tau}  &  =\frac{1}{2}y\left(  3\left(  1+x_{1}^{2}+x_{2}^{2}%
-y^{2}\right)  +\sqrt{6}\lambda x_{1}\right)  ,\label{ch.26}\\
\frac{d\mu_{1}}{d\tau}  &  =\sqrt{\frac{3}{2}}x_{1}\left(  4\kappa^{2}-\mu
_{1}^{2}\right)  \label{ch.27}%
\end{align}
where the variable $\mu_{1}$ is defined as $\mu_{1}=2\kappa\coth\left(
\kappa\phi_{1}\right)  $.

The stationary points~$\bar{P}=\left(  x_{1},y,x_{2},\mu_{1}\right)  $ of the
four-dimensional dynamical system (\ref{ch.24})-(\ref{ch.27}) which satisfy
the constraint condition (\ref{ch.20}) are%
\[
P_{1}^{\left(  \pm,\pm\right)  }=\left(  \pm1,0,0,\pm2\kappa\right)  ~,
\]%
\[
P_{2}^{\left(  \pm\right)  }=\left(  -\frac{\lambda}{\sqrt{6}},\sqrt
{1-\frac{\lambda^{2}}{6}},0,\pm2\kappa\right)  ~,
\]%
\[
P_{3}^{\left(  \pm\right)  }=\left(  0,0,\pm1,0\right)  ~,
\]%
\begin{align*}
P_{4}^{\left(  \pm\right)  }  &  =\left(  \frac{\sqrt{6}}{2\kappa-\lambda
},\sqrt{\frac{2\kappa}{2\kappa-\lambda}},\pm\sqrt{\frac{\lambda^{2}%
-2\lambda\kappa-6}{\left(  2\kappa-\lambda\right)  ^{2}}},-2\kappa\right)
~,~\\
P_{5}^{\left(  \pm\right)  }  &  =\left(  -\frac{\sqrt{6}}{2\kappa+\lambda
},\sqrt{\frac{2\kappa}{2\kappa+\lambda}},\pm\sqrt{\frac{\lambda^{2}%
+2\lambda\kappa-6}{\left(  2\kappa+\lambda\right)  ^{2}}},2\kappa\right)  .
\end{align*}

Points $P_{1}^{\left(  \pm,\pm\right)  },~~P_{2}^{\left(  \pm\right)  }$ have
the same physical properties as for $P_{1}^{\left(  \pm\right)  }$ and
$P_{2}$, for the quintessence model. In particular they can be seen as the
extension of the stationary points for the quintessence in the
four-dimensional manifold of the variables $\left(  x_{1},y,x_{2},\mu
_{1}\right)  $.

Points $P_{3}^{\left(  \pm\right)  }$ describe universes dominated by the
kinetic term of the second scalar field, while the parameter for the equation
of state for the effective fluid is $w_{eff}\left(  P_{3}^{\left(  \pm\right)
}\right)  =1$. Points $P_{4}^{\left(  \pm\right)  },\ $are real and physically
accepted when $\left\{  \kappa>0~,~\lambda\leq-\sqrt{6}\right\}  \cup\left\{
-\sqrt{6}<\lambda<0~,~2\kappa\geq\frac{\lambda^{2}-6}{\lambda}\right\}  $,
while points $P_{5}^{\left(  \pm\right)  }$ exist when $\left\{
\kappa>0~,~\lambda>\sqrt{6}\right\}  \cup\left\{  0<\lambda\leq\sqrt
{6}~,~2\kappa>\frac{6-\lambda^{2}}{\lambda}\right\}  $. The exact solutions at
the critical points describe scaling solutions with equation of state
parameter for the effective fluid%
\[
w_{eff}\left(  P_{4}^{\left(  \pm\right)  }\right)  =1+\frac{4\kappa}%
{\lambda-2\kappa}~,~w_{eff}\left(  P_{5}^{\left(  \pm\right)  }\right)
=1-\frac{4\kappa}{\lambda+2\kappa}~,~
\]
from were we can infer that $w_{eff}\left(  P_{4}^{\left(  \pm\right)
}\right)  <-\frac{1}{3}$ for $\left\{  \lambda\leq-\sqrt{2}~,~\kappa
>-\lambda\right\}  \cup\left\{  -\sqrt{2}<\lambda<0~,~2\kappa\geq\frac
{\lambda^{2}-6}{\lambda}\right\}  $ and $w_{eff}\left(  P_{5}^{\left(
\pm\right)  }\right)  <-\frac{1}{3}$ when $\left\{  \kappa>\lambda
~,~\lambda>\sqrt{2}\right\}  ~\cup~\left\{  0<\lambda<\sqrt{2}~,~2\kappa
\geq\frac{\lambda^{2}-6}{\lambda}\right\}  $. For the stability analysis we
apply the constraint equation (\ref{ch.20}) in order to reduce the dimension
of the dynamical system by one, in particular we replace $y=\sqrt{1-x_{1}%
^{2}-x_{2}^{2}}$.

The eigenvalues of the stationary points $P_{1}^{\left(  \pm,\pm\right)  }$
are
\[
e_{1}\left(  P_{1}^{\left(  \pm,\pm\right)  }\right)  =\varepsilon\sqrt
{6}\lambda+6~,~e_{2}\left(  P_{1}^{\left(  \pm,\pm\right)  }\right)
=-2\varepsilon\sqrt{6}\kappa~,~e_{3}\left(  P_{1}^{\left(  \pm,\pm\right)
}\right)  =-\varepsilon\sqrt{6}\kappa~,~\varepsilon=\left(  \pm1\right)
\left(  \pm1\right)  \text{.}%
\]
Hence, for $\varepsilon=+1$, the stationary points are attractors
when$~\kappa>0$ and $\lambda<-\sqrt{6}$, while for $\varepsilon=-1$, the
stationary points are attractors for $\lambda>\sqrt{6}$ and $\kappa<0$.

For the stationary points $P_{2}^{\left(  \pm\right)  }$ the eigenvalues are%
\[
e_{1}\left(  P_{2}^{\left(  \pm\right)  }\right)  =\frac{1}{2}\left(
\lambda^{2}-6\right)  ~,~e_{2}\left(  P_{2}^{\left(  \pm\right)  }\right)
=2\kappa\lambda~~,~e_{3}\left(  P_{2}^{\left(  \pm\right)  }\right)  =\frac
{1}{2}\left(  \lambda^{2}+2\kappa\lambda-6\right)
\]
which follows that $P_{2}^{\left(  \pm\right)  }$ are attractors when
$\left\{  -\sqrt{6}<\lambda<0~,~\kappa>0\right\}  \cup\left\{  0<\lambda
<\sqrt{6}~,~\kappa<0\right\}  $.

For the stationary points $P_{3}^{\left(  \pm\right)  }$ the stationary points
are%
\[
e_{1}\left(  P_{3}^{\left(  \pm\right)  }\right)  =6~,~e_{2}\left(
P_{3}^{\left(  \pm\right)  }\right)  =\sqrt{6}\kappa~\ ,~e_{3}\left(
P_{3}^{\left(  \pm\right)  }\right)  =-\sqrt{6}\kappa
\]
from where we infer that the points are always saddle points.

As far as the stationary points $P_{4}^{\left(  \pm\right)  }$ are concerned
it follows that one of the eigenvalues is $e_{1}\left(  P_{4}^{\left(
\pm\right)  }\right)  =\frac{12\kappa}{2\kappa-\lambda}$, which follows that
it is always positive when~$P_{4}^{\left(  \pm\right)  }$ are real points,
hence the exact solutions at these two points are always unstable. Similarly,
for $P_{5}^{\left(  \pm\right)  }$ one of the eigenvalues is derived to be
$e_{1}\left(  P_{5}^{\left(  \pm\right)  }\right)  =\frac{12\kappa}%
{2\kappa+\lambda}$ which is always positive hence the two exact solutions at
$P_{5}^{\left(  \pm\right)  }$ are always unstable.

We observe that the existence of the second scalar field introduce new
stationary points in the cosmological evolution which can describe various
cosmological eras. In particular, the additional points $~P_{3}^{\left(
\pm\right)  },~P_{4}^{\left(  \pm\right)  }$ and $~P_{5}^{\left(  \pm\right)
}$, where the second scalar field contributes, provide unstable exact solutions,
while points $P_{2}^{\left(  \pm\right)  }$ can be seen as the future
attractors. However, the set of points $P_{4}^{\left(  \pm\right)  }%
,~P_{5}^{\left(  \pm\right)  }$ can describe either a matter dominated era, or
the early acceleration phase. If $w_{eff}\left(  P_{4}^{\left(  \pm\right)
}\right)  =0$, such that the point to describe the matter epoch, it follows
$2\kappa=-\lambda$ and the coordinates of the points are%
\[
^{\left[  m\right]  }P_{4}^{\left(  \pm\right)  }=\left(  -\sqrt{\frac{3}{2}%
}\frac{1}{\lambda},\frac{1}{\sqrt{2}},\pm\frac{\sqrt{2\lambda^{2}-6}}%
{2\lambda},\lambda\right)
\]
with $\left\vert \lambda\right\vert >\sqrt{3}$. Similarly
$w_{eff}\left(  P_{5}^{\left(  \pm\right)  }\right)  =0$ gives $2\kappa
=-\lambda$, \ and in that case$~^{\left[  m\right]  }P_{4}^{\left(
\pm\right)  }=^{\left[  m\right]  }P_{5}^{\left(  \pm\right)  }$. 

We observe that in this two-scalar fields model we do not recover the results
presented in \cite{ns03,ns05} and that is because we used a different
reparametrization for the scalar fields, which means that the scalar fields
have different interaction terms. We continue our analysis by considering
three-scalar fields.

\subsection{Three-field model}

For the three-scalar field model, $\Phi^{A}=\left(  \phi_{1},\phi_{2},\phi
_{3}\right)  $, and, as before, the potential $V\left( \Phi^{A}\right)  =V_{0}e^{\lambda\phi_1}$ the
field equations in the dimensional variables are written%
\begin{align}
\frac{dx_{1}}{d\tau}  &  =\frac{1}{2}\left(  3x_{1}^{3}-3x_{1}\left(
1+y^{2}-x_{2}^{2}\right)  -\sqrt{6}\lambda y^{2}+\sqrt{6}\mu_{1}\left(
x_{2}^{2}+x_{3}^{2}\right)  \right)  ,\label{ch.28}\\
\frac{dx_{2}}{d\tau}  &  =\frac{1}{2}x_{2}\left(  3\left(  x_{1}^{2}+x_{2}%
^{2}+x_{3}^{2}-y^{2}-1\right)  -\sqrt{6}\mu_{1}x_{1}\right)  +\frac{\sqrt{6}%
}{2}\mu_{2}x_{3}^{2},\label{ch.29}\\
\frac{dx_{3}}{d\tau}  &  =\frac{1}{2}x_{3}\left(  3\left(  x_{1}^{2}+x_{2}%
^{2}+x_{3}^{2}-y^{2}-1\right)  -\sqrt{6}\left(  \mu_{1}x+\mu_{2}x_{2}\right)
\right) \label{ch.30}\\
\frac{dy}{d\tau}  &  =\frac{1}{2}y\left(  3\left(  1+x_{1}^{2}+x_{2}^{2}%
+x_{3}^{2}-y^{2}\right)  +\sqrt{6}\lambda x_{1}\right)  ,\label{ch.31}\\
\frac{d\mu_{1}}{d\tau}  &  =\sqrt{\frac{3}{2}}x_{1}\left(  4\kappa^{2}-\mu
_{1}^{2}\right)  ,\label{ch.32}\\
\frac{d\mu_{2}}{d\tau}  &  =\sqrt{\frac{3}{2}}\left(  4\kappa^{2}x_{2}-\mu
_{1}\mu_{2}x_{1}-x_{2}\left(  \mu_{1}^{2}+3\mu_{2}\right)  +\sqrt{\frac{3}{2}%
}\frac{\left(  \mu_{1}\mu_{2}\right)  ^{2}}{4\kappa^{2}}x_{2}\right)  ,
\label{ch.33}%
\end{align}
where $\mu_{2}=2\kappa\frac{\cot\kappa\phi_{2}}{\sinh\kappa\phi_{1}}$. The
stationary points $P=\left(  x_{1},y,x_{2},\mu_{1},x_{3},\mu_{2}\right)  $ for
the latter dynamical system are%
\[
P_{1}^{\left(  \pm,\pm\right)  }=\left(  \pm1,0,0,\pm2\kappa,0,0\right)  ,
\]%
\[
P_{2}^{\left(  \pm\right)  }=\left(  -\frac{\lambda}{\sqrt{6}},\sqrt
{1-\frac{\lambda^{2}}{6}},0,\pm2\kappa,0,0\right)  ,
\]%
\[
P_{3}^{\left(  \pm,\pm\right)  }=\left(  0,0,\pm1,0,0,\pm\frac{2\sqrt{3}%
\kappa}{3}\right)  ,~
\]%
\[
P_{4}^{\left(  \pm,\pm\right)  }=\left(  \frac{\sqrt{6}}{2\kappa-\lambda
},\sqrt{\frac{2\kappa}{2\kappa-\lambda}},\pm\sqrt{\frac{\lambda^{2}%
-2\lambda\kappa-6}{\left(  2\kappa-\lambda\right)  ^{2}}},-2\kappa,0,\pm
\frac{\sqrt{6}\kappa}{\sqrt{\lambda^{2}+2\kappa\lambda-6}}\right)  ,
\]%
\[
P_{5}^{\left(  \pm,\pm\right)  }=\left(  -\frac{\sqrt{6}}{2\kappa+\lambda
},\sqrt{\frac{2\kappa}{2\kappa+\lambda}},\pm\sqrt{\frac{\lambda^{2}%
+2\lambda\kappa-6}{\left(  2\kappa+\lambda\right)  ^{2}}},2\kappa,0,\pm
\frac{\sqrt{6}\kappa}{\sqrt{\lambda^{2}+2\kappa\lambda-6}}\right)  ,
\]%
\[
P_{6}^{\left(  \pm\right)  }=\left(  0,0,0,0,\pm1,0\right)  ,
\]%
\[
P_{7}^{\left(  \pm\right)  }=\left(  \frac{\sqrt{6}}{2\kappa-\lambda}%
,\sqrt{\frac{2\kappa}{2\kappa-\lambda}},x_{2},-2\kappa,\pm\sqrt{\frac{\left(
\lambda-2\kappa\right)  \left(  \lambda-x_{2}\left(  \lambda-2\kappa\right)
\right)  -6}{\left(  \lambda-2\kappa\right)  ^{2}}},0\right)  ,
\]%
\[
P_{8}^{\left(  \pm\right)  }=\left(  -\frac{\sqrt{6}}{2\kappa-\lambda}%
,\sqrt{\frac{2\kappa}{2\kappa-\lambda}},x_{2},2\kappa,\pm\sqrt{\frac{\left(
\lambda-2\kappa\right)  \left(  \lambda-x_{2}\left(  \lambda-2\kappa\right)
\right)  -6}{\left(  \lambda-2\kappa\right)  ^{2}}},0\right)  .
\]

Points $P_{1}^{\left(  \pm,\pm\right)  },~P_{2}^{\left(  \pm\right)  }%
,~P_{3}^{\left(  \pm,\pm\right)  },~$ are the extensions of the stationary
points for the two fields in the six-dimensional space, while the new
stationary points are $P_{4}^{\left(  \pm,\pm\right)  },P_{5}^{\left(  \pm
,\pm\right)  },~P_{6}^{\left(  \pm\right)  },~P_{7}^{\left(  \pm\right)  }$
and $P_{8}^{\left(  \pm\right)  }$. Points $P_{7}^{\left(  \pm\right)
},~P_{8}^{\left(  \pm\right)  }$~are actually surfaces in the space $\left\{
x_{2},x_{3}\right\}  ,$while when $x_{2}=\pm\sqrt{\frac{\lambda^{2}%
-2\lambda\kappa-6}{\left(  2\kappa-\lambda\right)  ^{2}}}$ the points reduce
to that of $P_{4}^{\left(  \pm\right)  }$,~$P_{5}^{\left(  \pm\right)  }$ for
the two-scalar fields model.

The physical properties of points~$P_{1}^{\left(  \pm,\pm\right)  }%
,~P_{2}^{\left(  \pm\right)  },~P_{3}^{\left(  \pm,\pm\right)  }$ are the same
as for the two fields case; $P_{6}^{\left(  \pm\right)  }$ describe universes
dominated by the kinetic part for the second scalar field, and the parameter
for the equation of state is calculated $w_{eff}\left(  P_{6}^{\left(
\pm\right)  }\right)  =1$. Furthermore, the physical solutions at the
stationary points $P_{4}^{\left(  \pm,\pm\right)  },P_{5}^{\left(  \pm
,\pm\right)  },~P_{7}^{\left(  \pm\right)  }$ and $P_{8}^{\left(  \pm\right)
}$ are the same with that of points $P_{4}^{\left(  \pm\right)  }$%
and~$P_{5}^{\left(  \pm\right)  }$.

Consequently, the introduction of the new scalar field provides families of
new stationary points but with not new physical properties. We proceed with
the discussion on the stability of the exact solutions at the critical points.

As far as the stability properties of the points $P_{1}^{\left(  \pm
,\pm\right)  },~P_{2}^{\left(  \pm\right)  },~P_{3}^{\left(  \pm,\pm\right)
}$ are concerned, points $P_{3}^{\left(  \pm,\pm\right)  }$ are always saddle
points, while $P_{1}^{\left(  \pm,\pm\right)  },~P_{2}^{\left(  \pm\right)  }$
can be attractors for the same values of the free parameters as for the
two-scalar field model. The eigenvalues of the linearized system around the
points $P_{6}^{\left(  \pm\right)  }$ are derived%
\[
e_{1}\left(  P_{6}^{\left(  \pm\right)  }\right)  =\frac{3}{2}~,~e_{2,3}%
\left(  P_{6}^{\left(  \pm\right)  }\right)  =-\frac{3+\sqrt{3\left(
3+32\kappa^{2}\right)  }}{4}~,~e_{4,5}\left(  P_{6}^{\left(  \pm\right)
}\right)  =-\frac{3-\sqrt{3\left(  3+32\kappa^{2}\right)  }}{4},
\]
from we conclude that the points are saddle. Finally, for the rest sets of
points, namely, $P_{4}^{\left(  \pm,\pm\right)  },P_{5}^{\left(  \pm
,\pm\right)  },~P_{7}^{\left(  \pm\right)  }$ and $P_{8}^{\left(  \pm\right)
}$ at least of one of the eigenvalues has the value $e_{1}\left(
P_{4}^{\left(  \pm\right)  }\right)  =\frac{12\kappa}{2\kappa-\lambda}$ or
$e_{1}\left(  P_{5}^{\left(  \pm\right)  }\right)  =\frac{12\kappa}%
{2\kappa+\lambda}$ from we can infer that the exact solutions at the
stationary points are always unstable, while the points are found to be
saddle. Therefore the only future attractors can be the exact solutions of the
two-scalar fields model, namely points $P_{1}^{\left(  \pm,\pm\right)  }%
,~P_{2}^{\left(  \pm\right)  }$.

\subsection{$N$-field model}

From the given analysis we see that the two-scalar field models provide
new physical solutions in comparison with the quintessence model. On the other
hand, in the three-scalar field model the new stationary points do not
describe new physical universes, while the only attractors are the exact
solution of the quintessence field. \ In order to understand that let us study
the invariant surfaces of the multi-field model.

The point-like Lagrangian of the $N-$fields model reads%
\begin{equation}
L\left(  a,\dot{a},\Phi,\dot{\Phi}\right)  =-3a\dot{a}^{2}+\frac{1}{2}%
a^{3}\left(  \dot{\phi}_{1}^{2}+\sinh^{2}\left(  \kappa\phi_{1}\right)
\left(  \dot{\phi}_{2}^{2}+\sin^{2}\left(  \kappa\phi_{2}\right)  \left(
\dot{\phi}_{3}^{2}+...\right)  \right)  \right)  -V_{0}a^{3}e^{\lambda\phi
_{1}}.\label{lan01}%
\end{equation}
The latter point-like Lagrangian function is autonomous which means that
admits as Noether symmetry \cite{noo1} the vector field $\partial_{t}$ with
conservation law the Hamiltonian function $\mathcal{H}\left(  a,\dot{a}%
,\Phi,\dot{\Phi}\right)  =h$, which is nothing else than the constraint
equation (\ref{ch.17}) which means that $\mathcal{H}\left(  a,\dot{a}%
,\Phi,\dot{\Phi}\right)  =0$. Hence it follows%
\begin{equation}
\mathcal{H}\left(  a,\dot{a},\Phi,\dot{\Phi}\right)  \equiv-3a\dot{a}%
^{2}+\frac{1}{2}a^{3}\left(  \dot{\phi}_{1}^{2}+\sinh^{2}\left(  \kappa
\phi_{1}\right)  \left(  \dot{\phi}_{2}^{2}+\sin^{2}\left(  \kappa\phi
_{2}\right)  \left(  \dot{\phi}_{3}^{2}+...\right)  \right)  \right)
+V_{0}a^{3}e^{\lambda\phi_{1}}=0\label{lan02}%
\end{equation}

However, except from the Hamiltonian function, the dynamical system described
by the point-like Lagrangian (\ref{lan01}) admits additional conservation
laws. Someone can apply Noether's theorem or other methods to construct
conservation laws \cite{noo2}. Indeed, one of conservation laws admitted by
the dynamical system is the quadratic conservation law \cite{noo3}%
\begin{equation}
I_{0}^{2}=a^{6}\sinh^{4}\left(  \kappa\phi_{1}\right)  \left(  \dot{\phi}%
_{2}^{2}+\sin^{2}\left(  \kappa\phi_{2}\right)  \left(  \dot{\phi}_{3}%
^{2}+...\right)  \right)  \label{lan03}%
\end{equation}
where now the Hamiltonian function reads%
\begin{equation}
\mathcal{H}\left(  a,\dot{a},\Phi,\dot{\Phi}\right)  \equiv-3a\dot{a}%
^{2}+\frac{1}{2}a^{3}\left(  \dot{\phi}_{1}^{2}+V_{0}a^{3}e^{\lambda\phi_{1}%
}\right)  +\frac{I_{0}^{2}}{2a^{3}\sinh^{2}\left(  \kappa\phi_{1}\right)  }.
\end{equation}

For $I_{0}=0$, the dynamics reduce to that of quintessence field, however for
$I_{0}\neq0$ the new term drives the dynamics and provide the new additional
physical behaviour. The quadratic conservation law $I_{0}$ defines an
invariant surface of the dynamical system, while we observe that invariant
sub-surfaces exist also in $I_{0}$, however they do not affect the final
dynamics. \

Consequently, the following theorem.

\textbf{Theorem: }\textit{For the Chiral model with }$N-$\textit{fields, with
}$N>2$\textit{ the exact cosmological solutions which are described\ by the
stationary points have the same physical properties of the two-scalar field
model,~\thinspace\thinspace}$N=2$\textit{, which means that the possible
solutions are: the stiff fluid solution, scaling solution of quintessence and
a scaling solution where two or more scalar fields contributes in the total
cosmological fluid. Finally, the quintessence model is the future attractor of
the }$N-$\textit{field Chiral model. }

The above theorem says that in this Chiral theory, the consideration of more
than two scalar fields is only of mathematical interests and there is not any
new physical properties. The scalar fields are nonlinear
and there are interaction terms which follow from the kinetic terms of the
Lagrangian function. Of course the consideration of a more general potential
function can lead to a different result, but again the limit of the
quintessence field will always exists, and the above theorem will holds.

\section{Conclusions}

\label{sec5}

In this work we considered a multi-field  cosmological model. Specifically,  we consider the existence of $N$-fields in a spatially flat FLRW background space
which interact in the kinetic terms. That model is the multi-field
extension of the Chiral theory, where the dynamics of the fields are over an
space of constant curvature.

We focus on the contribution of the new fields in the cosmological
evolution.\ In particular we wanted to answer  the question of how essential
is the introduction of the new scalar fields in the cosmological dynamics. In
order to work in this direction we focus on the asymptotic analysis for the
dynamics of the cosmological field equations. \ We presented a detailed
analysis of the stationary points for the one-field model, which is the
quintessence with exponential scalar field, for the two-fields Chiral model
and for the three-fields extension. From these results we found that when we
pass from the one-field to the two-fields model, the cosmological behaviour
becomes richer. New stationary points follows which provide new physical
solutions.

However, when we pass from the two-fields to the three-fields model, the new
stationary points do not provide new physical solutions, while the additional
stationary points are only of mathematical interests. For the $N-$fields model,
and with the use of invariant functions, we describe the physical properties, and
the asymptotic behaviour for the field equations in a theorem. The theorem
states that the only possible physical solutions of the $N$-fields model that exists are those that describes: stiff fluid dominated universes, the quintessence scaling solution and a scaling solution where two or more scalar fields contributes in the total
cosmological fluid.

From the above analysis we conclude that in this specific theory, the
consideration of more that two scalar fields does not affect the physics.

\begin{acknowledgments}
AP \& GL were funded by Agencia Nacional de Investigaci\'{o}n y Desarrollo -
ANID through the program FONDECYT Iniciaci\'{o}n grant no. 11180126.
Additionally, GL is supported by Vicerrector\'{\i}a de Investigaci\'{o}n y
Desarrollo Tecnol\'{o}gico at Universidad Catolica del Norte.
\end{acknowledgments}

\end{document}